\documentclass[11pt]{article}
\usepackage{amsfonts}
\usepackage{bm}
\usepackage[margin=1.4in]{geometry}
\usepackage[affil-it]{authblk}
\usepackage{graphicx}
\usepackage{floatrow}
\usepackage{subcaption}
\usepackage{bmpsize}
\usepackage{epstopdf}
\usepackage{amsmath}
\usepackage{xcolor}
\usepackage[toc,page]{appendix}
\usepackage{booktabs}
\usepackage{hyperref}

\renewcommand\appendix{\par
	\setcounter{section}{0}%
	\setcounter{subsection}{0}%
	\setcounter{table}{0}
	\setcounter{figure}{0}
	\gdef\varthetable{\Alph{table}}
	\gdef\thefigure{\Alph{figure}}
	\section*{Appendix}
	\gdef\thesection{\Alph{section}}
	\setcounter{section}{1}}
\begin{document}
\title{Scattering from a non-linear structured interface}
\author[1,2]{Domenico Tallarico}
\author[2]{Natalia V. Movchan}
\author[2]{Alexander B. Movchan}

\affil[1]{EMPA, Laboratory for Acoustics/ Noise Control, \"{U}berlandstrasse 129, 8600 D\"{u}bendorf, Switzerland} 
\affil[2]{Department of Mathematical 
	Sciences,
	Mathematical Sciences Building,The University of Liverpool, L69 7ZL, Liverpool, United Kingdom}
\date{}
\maketitle

\vspace{.1in}

\centerline{\it In honour of Professor Davide Bigoni on the occasion of his 60th Birthday}

\vspace{.1in}

\begin{abstract}
We review the scattering from non-linear interfaces containing buckling elastic beams. An illustrative example is discussed here of  scattering of linear elastic pressure waves from a two-mass system connected by a non-linear structured interface modelled as  elastica. In the first instance, the  interaction between the masses is linearised.
This allows for the study of a time-harmonic transmission model problem in 
the subcritical regime. Subsequently, we consider the transient problem associated with a non-linear ineraction within the interface.
The effect of  non-linearity is shown to suppress the transmission resonance observed in the linearised formulation. 
\end{abstract}

 \section{Introduction}
  
  The idea of longitudinal and transverse waves interaction in beams, in the framework of the configurational forces approach, was introduced and developed by Bigoni {\em et al.} \cite{Bigoni2012_PRSA,Bigoni2014_JMPS,Bigoni2015_MM, Bigoni2014_PRSA}
  The results of these papers were applied, in particular, to modelling of locomotion in challenging biological applications. 
  The variational approach of Bigoni {\em et al.} \cite{Bigoni2012_PRSA,Bigoni2014_JMPS, Bigoni2015_MM, Bigoni2014_PRSA}  has led to an accurate evaluation of the longitudinal configurational force, which exists within an elastic beam subjected to bending  by a moment.
  In the recent paper by Bosi {\em et al.} \cite{Bosi2016_InPress}, the solution of the elastica and  stability analysis in a dynamic perturbation problem were developed in the context of an asymptotic self-stabilisation of a continuous elastic structure. 
  
The Bigoni's approach based on the notion of configurational forces, which connect the bending or torsional moments and the longitudinal force in the deformed beam, has lead to the rigorous modelling and deep physical understanding of  instability of constrained elastic beams under transverse load or torsion in a range of important applications.  In particular, the work    \cite{Bigoni2014_JMPS} considers 
a  blade, which is forced into an elastic movable junction: bifurcation and stability analysis  has revealed several unusual features, which include the effect of restabilisation of the straight configuration, and the configurational forces are shown to be essential in the bifurcation analysis.  Dal Corso  \emph{et al.} in \cite{dal2017serpentine} have analysed a bio-inspired model of serpentine locomotion based on the release of kinetic energy, provided by the configurational potential energy stored in a beam while constrained by a curved channel. Furthermore, the general analysis of an  asymptotic self-restabilization of continuous elastic systems was presented  in \cite{Bosi2016_InPress}.  A structure that self-restabilizes is capable of restoring its initial shape after a post-bifurcation deformation. The problem has been solved in both quasi-static and dynamic frameworks, and it has been demonstrated that the asymptotic restabilisation is determined by the effects produced by configurational forces.  
Systematic analysis of material instability is included in the comprehensive research monograph \cite{bigoni2012nonlinear}. In particular, it includes elegant examples of bifurcation of a self-intersecting elastica as well as examples of bifurcations for structures under tensile loads. Restabilisation was also explained by analysing structures with  trivial configurations unstable at a certain load, and  later returning to a stable state  at a higher load.   

Non-local structured interface, both in statics and dynamics, were analysed by Bigoni and Movchan \cite{bigmov2002statics}, where transmission conditions were derived in conjunction with the analysis of connections between the discrete and continuous systems. In particular, the role of discrete structures was studied in the context of wave propagation in periodic systems, which included discrete and continuous constituents. The paper \cite{2013BigGuMoBr} has brought an idea of coupling between shear and pressure modes by structured interfaces incorporating rotations - a periodic system of inclusions was embedded into an elastic continuum via a finite thickness structured interface with tilted elements. Such structures were shown to be useful in control of negative refraction of elastic waves across two-dimensional structured interfaces. In the recent article \cite{tallarico2017edge} geometrically chiral lattice interfaces were incorporated into finite coatings surrounding crack-like defects; the dynamics of such interfaces has provided new insight into controlling dynamic response of multi-scale lattice systems, and shielding defects, which otherwise would be stress concentrators leading to possible structural failure.  

 In some elastic systems, physical non-linearities or geometrical factors may lead to the dynamic response of solids which cannot be treated in the standard framework of linear time-harmonic analysis. Research monographs by Achenbach \cite{achenbach2012wave} (Chapter 10) and Samsonov \cite{samsonov2001strain} give a comprehensive theoretical description and experimental survey of non-linear waves in solids, including localisation and well-posedness. Using a Lagrangian approach, Samsonov \cite{samsonov2001strain} derived double-dispersive wave equations in elastic solids, including associated conservation laws and solitary solutions. It has been shown that the double-dispersive wave equation is a continuous limit of some non-linear periodic elastic chains. Moreover, the double-dispersive wave equation captures non-linear wave phenomena in homogeneous and non-homogeneous elastic waveguides. Most notably, the amplitude of a solitary elastic deformation is focused by a tapered waveguides, \emph{i.e.} it is concentrated in a shorter interval of time past the tapered region, while it remains almost unchanged after travelling through randomly corroded regions. These phenomena bring 
 additional inspirations to the researchers working on 
 wave phenomena in structured solids and metamaterials, such as focusing via negative refraction and cloaking. 
  
The recent papers by Maurin and Spadoni \cite{Maurin2014_WM,Maurin2014_JSV,Maurin2016_WM1,Maurin2016_WM2} have discussed non-linear elastic waves, which may form in a periodic system of buckled beams. Conventionally, for an isotropic homogeneous elastic beam, in the leading-order linearised approximation, the flexural vibrations decouple from the longitudinal ones.
However, when the compressive longitudinal load reaches a critical value, an elastic beam buckles, and a periodic system of many buckled beams can be considered as a structured medium supporting dynamic localisation. 
In particular, a fourth-order homogenisation model was used by Maurin and Spadoni to describe dispersive properties of an infinite system of buckled beams connecting point masses periodically distributed within the array.  In traditional engineering designs, the post-bifurcation behaviour is often avoided, but as demonstrated in \cite{Maurin2014_WM,Maurin2014_JSV,Maurin2016_WM1,Maurin2016_WM2} such problems bring  interesting mathematical formulations, and some waveforms can be described as closed form solutions for the differential equation of the Boussinesq type. 
In the long-wave regime the homogenisation model is an acceptable way forward to describe a structured waveguide.  
  

In Fig. \ref{fig:f-nl} below, we show the force versus displacement for a structured interface. One may get an impression that the axial load transmitted by the structured interface is similar to the constitutive response of an elastic-perfectly plastic rod.  We refer to the classical publications on dynamic plasticity \cite{von1950propagation,bell1968large,kuscher1986non} and claim that on the one hand there is formal analogy between the elastic-perfectly plastic rod and the problem considered in this paper. On the other hand, our formulation models different problems describing waves in elastic-plastic media. In our case, the emphasis is on the use of configurational forces connecting the flexural displacement of an elastica with the longitudinal forces of the adiacent elastic rod, in the context of the dynamic response of a structured interface. 

The purpose of the present paper is to highlight new features in the dynamic response of elastic solids, which contain non-linear interfaces. The example shown here, demonstrates  the 
pre-bucked and post-buckled regimes for a system of elastic waveguides with masses connected by an elastic beam within the interface. The model  includes two semi-infinite linearly elastic rods separated by a buckling beam. The assumptions of the model can be extended to other physical configurations which may include a collection of flexural beams which are able to buckle at different critical loads. In the current model, it assumed that the semi-infinite elastic rods at both sides of the interface transmit longitudinal linear waves only.

The structure of the paper is as follows. In section \ref{sec:gov} we present the non-linear interface modelled as a two-mass system connected by an Euler-type beam which can buckle. The pre-buckling and post-buckling regimes are reviewed by standard Euler-type buckling techniques. The non-linear force-compression relations are given in implicit form for a general compression-level of the two-mass system, and as a truncated series expansion around a special value of the compression level.  The two-mass non-linear system is embedded into an homogeneous elastic bar which supports longitudinal waves. In section \ref{sec:th-scattering}, the solution to a scattering of a time-harmonic longitudinal wave  is discussed and transmission resonances are identified. A transient problem for waves reflecting from a non-linear interface is analysed in section \ref{transient}. Section \ref{conclusion} includes concluding remarks and discussion. 

   \section{Governing equations\label{sec:gov}}
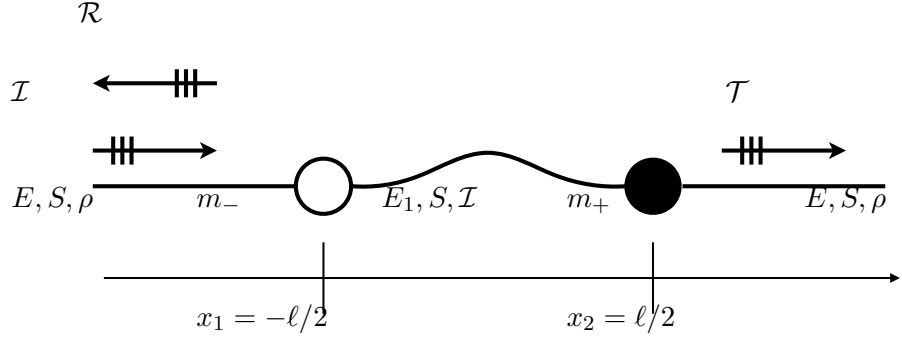
\begin{figure}[t!]
\centering
\includegraphics[trim={3cm 9cm  3cm 9cm},width=0.8\textwidth]{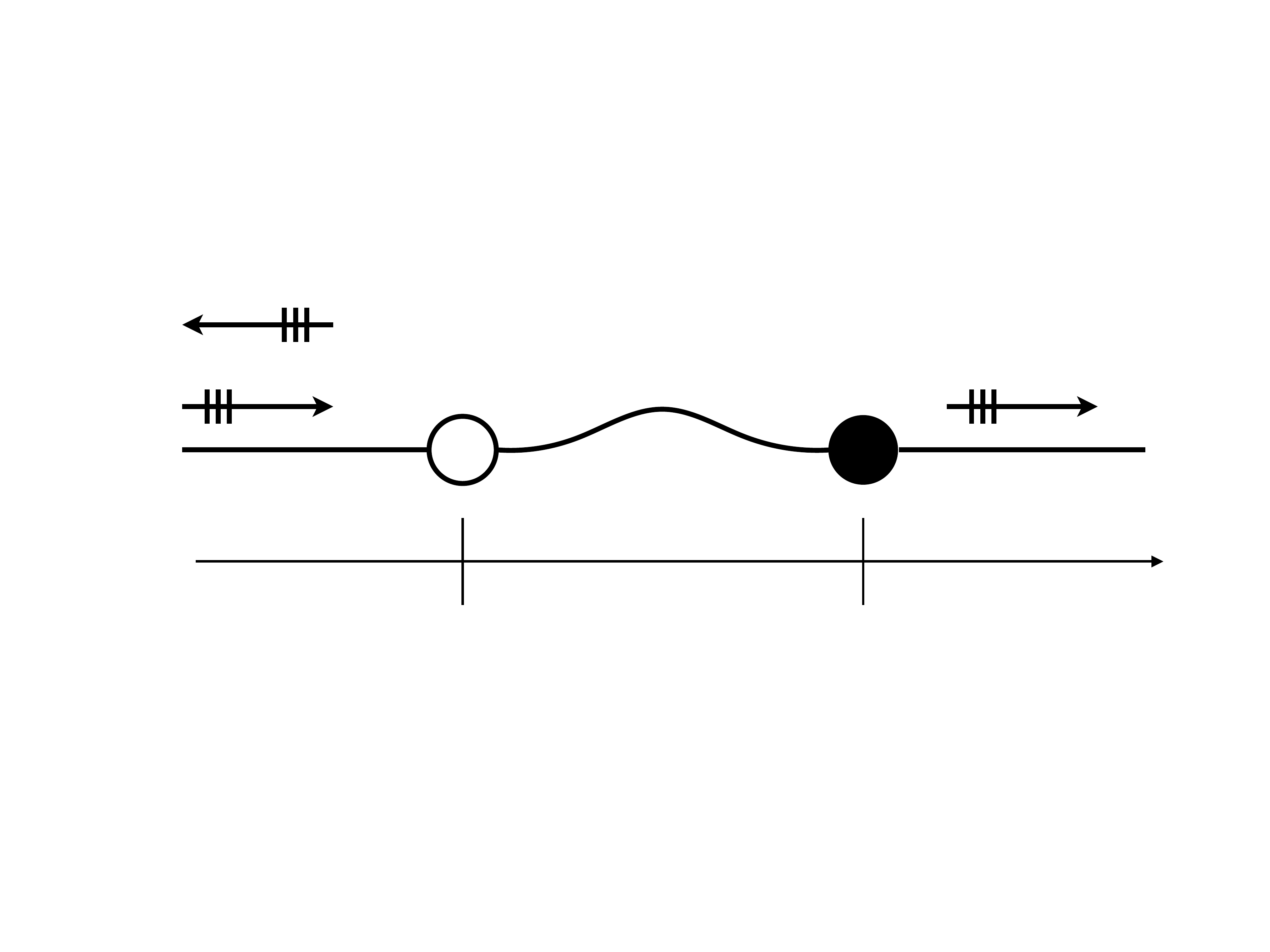}
		\begin{picture}(0,0)(0,0)
		\put(-270,50) {$m_-$}
		\put(-130,50) {$m_+$}
		\put(-270,5) {$x_1=-\ell/2$}
		\put(-130,5) {$x_2=\ell/2$}
		\put(-340,90){${\cal I}$}
		\put(-315,120){${\cal R}$}
		\put(-70,90){${\cal T}$}
		\put(-340,50){$E,S,\rho$}
		\put(-40,50){$E,S,\rho$}
		\put(-200,50){$E_1,S,{\cal I}$}
		\end{picture}
\caption{\label{fig:geom} Two semi-infinite longitudinal rods (straight lines), of cross-section $S$, Young's modulus $E$ and mass density $\rho$, are connected by a structured interface. The interface comprises  an    elastica arm  (cross-section $S$, second moment of intertia $\cal I$ and Young's modulus $E_1$) with masses $m_-$ (white circle) and $m_+$ (black circle) at the ends $x=-\ell/2$ and $x=+\ell/2$, respectively. }
\end{figure}

  Consider an infinite rod of cross section $S$, Young's modulus $E$ and density $\rho$, where only longitudinal waves can propagate. Let us assume that a section of length   $\ell$   is  removed and replaced with a massless flexural beam of bending stiffness ${\cal B} = E_1 {\cal I}$, and horizontal in its undeformed configuration. Here, $E_1$ is the Young's modulus of the beam, and ${\cal I}$ is the second moment of inertia of its cross-section, of the same area $S$ as the exterior elastic rod. The structured interface is formed by point masses, $m_-$  and $m_+$  attached to the ends of the  flexural beam, as shown in Fig. \ref{fig:geom}.  \subsection{Equations of motion}
    The displacement of the ends of the massless flexural beam are denoted by 
    $u_{\pm}(t)=
    u(\pm\ell/2,t),$
    and $u(x,t)$ satisfies
  \begin{equation}\label{eq:pde-longitudinal}
  \frac{\partial^2}{\partial x^2}u(x,t)-\frac{1}{v^2} \frac{\partial^2}{\partial t^2}u(x,t)=0, {\rm \,\,\,\,for\,\,\,\,} {\|x\|>\ell/2},
  \end{equation}
  with $v^{2}=E/\rho$.

   \subsection{Transmission conditions}
 The equations of motion for each of the masses $m_+$ and $m_-$ have the form 
 \begin{equation}\label{eq:bc-derivative}
 \left.\frac{\partial}{\partial x}u(x,t)\right|_{x=\pm\ell/2}=\frac{1}{ E  S}\left( \mp m_{\pm} \frac{{\rm d}^2}{{\rm d}t^2}u_\pm+{\cal F}(u_+,u_-)\right), 
 \end{equation}
 where 
 ${\cal F}$ represents the force describing a non-linear interaction across the interface, as described in  section \ref{sec:nl-int} below. 
 We also use the notation
 $$
 F(\chi) = {\cal F}(u_+,u_-), ~\mbox{where} ~ \chi = (u_--u_+)/{\ell}. 
 $$
  The continuity of displacements at $x=\pm \ell/2$ implies 
 \begin{equation}\label{eq:bc-continuity}
 \left.u(x,t)\right|_{x=\pm\ell/2}=u_\pm(t).
 \end{equation}
  Eqs \eqref{eq:bc-derivative} and \eqref{eq:bc-continuity} form the transmission conditions for the structured interface. 
 \subsection{Non-linear interaction \label{sec:nl-int}}
In this section, we discuss the exact load-displacement relation of a clamped-clamped Euler-type elastica. This also includes approximate load-displacement relations for small post-buckling displacement. 
 \begin{figure}[h!]
\centering
\includegraphics[width=0.8\textwidth]{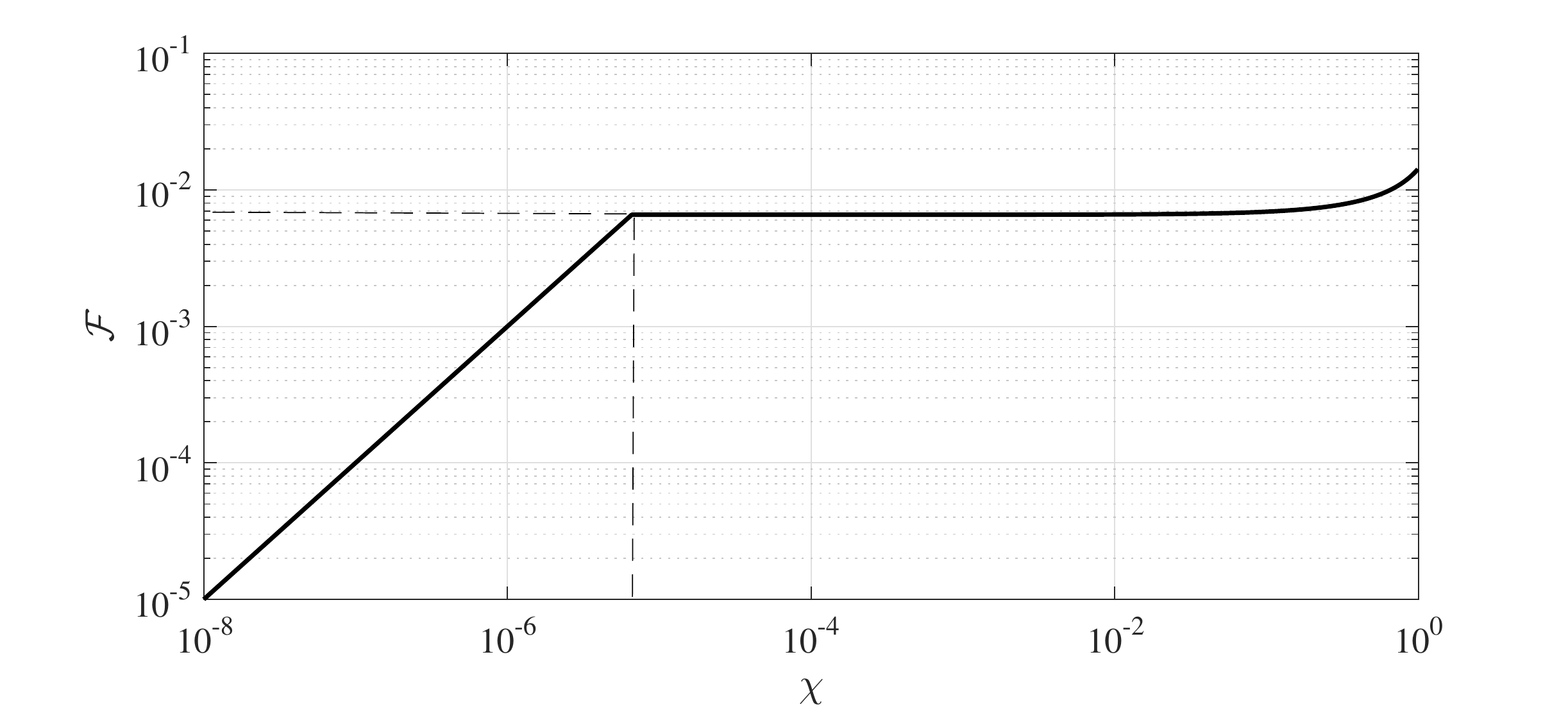}
		\begin{picture}(0,0)(0,0)
		\put(-285,115)  {$P_c$}
		\put(-225,30) {$\chi_{\rm cr}$}
		\end{picture}
\caption{\label{fig:f-nl} Axial load $\cal F$ as a function of the normalised axial compression $\chi=|u_--u_+|/\ell$, in the pre-buckling regime  ($\chi<\chi_{\rm cr}$, see Eq. \eqref{eq:f-prebuck}) and post-buckling regime  ($\chi>\chi_{\rm cr}$, see Eq. \eqref{eq:load-displacement}).  We fix $P_c=6.6\times10^{-3}$ and $\chi_{\rm cr}=6.6\times10^{-6}$.  }
\end{figure}
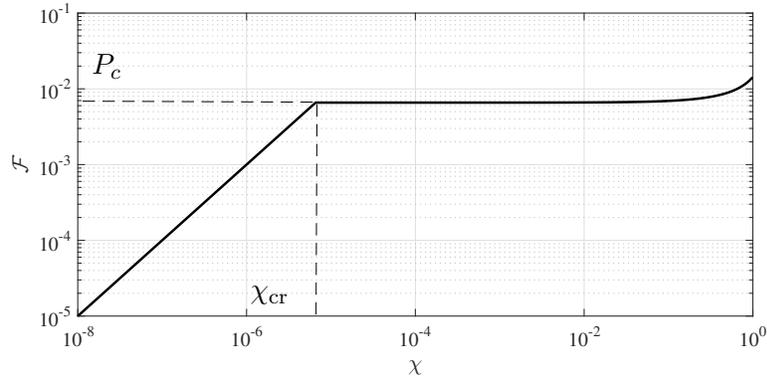
The main focus  is on the non-linear interaction across the interface, for which the  exact load-displacement relation is given by the solution of the following system of non-linear equations 
\begin{equation}\label{eq:load-displacement}
\begin{cases}
\ell \sqrt{F/B}  = 4 {\rm K}(c)\\
\\
\chi - \chi_{\rm{cr}} =  2 (1-{{\rm E}(c)}/{{\rm K}(c)})
\end{cases}
,
\end{equation}
where $B$ is the bending stiffness, $\chi_{\rm{cr}}$ is the critical compression, and  
\begin{equation}\label{eq:def_cei}
{\rm K}(c) = \int_0^{\pi/2} \frac{{\rm d}\phi}{\sqrt{1-c^2 \sin^2 \phi}},~~~{\rm and}~~~{\rm E}(c) = \int_0^{\pi/2} \sqrt{1-c^2 \sin^2 \phi}~ {\rm d} \phi
\end{equation}
are complete elliptic integrals of the first and second type, respectively. 
The force ${\cal F}$ as a function of the axial strain $\chi$ is shown in Fig. \ref{fig:f-nl}.  
Here, we use the following parameters
\begin{equation}
c^2=\sin^2(\vartheta^*/2),~~~\ell_x / \ell  = 1 - \chi, ~~{\rm with}~~\chi = (u_--u_+)/\ell=\chi_{\rm cr} + \Delta \chi ,\texttt{\texttt{\texttt{}}}
\end{equation} 
where $\vartheta^*$ is the angle of rotation of the tangent line at  inflection points (zero moments) along the elastica.
Further down, in Section \ref{Sec232}, we also discuss approximate  relations for the load-displacement equation, which  
are obtained  
for $0<\Delta \chi\ll1$, \emph{i.e.} 
for small post-buckling compression. 
We assume that the elastica undergoes a simple Euler-type buckling, and 
do not consider 
higher-order buckling modes.

\subsubsection{Discussion of the non-linear interaction based on the Euler elastica model}
Equation (\ref{eq:load-displacement})  follows from the classical  theory  
 of stability of structures, discussed in detail in the monographs   \cite{bigoni2012nonlinear, cedolin2010stability,thompson1973general}. For convenience of the reader we discuss the derivation of the formula (\ref{eq:load-displacement}). We note that the case of hinge junctions is discussed fully  in \cite{cedolin2010stability}; here we apply a similar scheme to the case of a movable clamp.  The rotation $\vartheta(s)$ at a given position $s$ along the deformed configuration of the elastica arm is governed by the second-order differential equation
\begin{equation}\label{eq:ode-elastica}
E_1 {\cal I} \vartheta^{\prime\prime} + F \sin \vartheta =  0,
\end{equation}
where $\cal{I}$ is the second moment of inertia,  $E_1$ is the Young's modulus of the beam within the interface, and $B = E_1 \cal{I} $ is the flexural stiffness. 
Multiplying the left hand side of Eq. \eqref{eq:ode-elastica} by $\vartheta^\prime$ we deduce 
\begin{align}\label{eq:ode-elastica-1}
\vartheta^\prime\left( E_1 {\cal I} \vartheta^{\prime\prime} + 2 F \sin \frac{\vartheta}{2} \cos \frac{\vartheta}{2} \right) 
=2 {\cal F} 
\frac{\rm d}{{\rm d} s} \left(\frac{{\vartheta^\prime}^2}{4 \lambda^2} + \sin^2 \frac{\vartheta}{2} \right)=0 ,\end{align}
and hence
\begin{equation}
\frac{{\vartheta^\prime}^2}{4 \lambda^2} + \sin^2 \frac{\vartheta}{2}= c^2, \label{eq:ode-elastica-1a}
\end{equation}
with $c^2$ being a constant of integration and $\lambda^2 = F / B$.  
Using (\ref{eq:ode-elastica-1a}), together with the transmission conditions at the interface boundary, we derive  
\begin{equation}
c^2 = \frac{{\vartheta^\prime}^2(0)}{4 \lambda^2}=\sin^2 \frac{\vartheta(\ell^*)}{2} = \frac{{\vartheta^\prime}^2(\ell/2)}{4 \lambda^2},
\end{equation} 
where  $\ell^*$ denotes the position of   the inflexion point, \emph{i.e.} the coordinate of zero  moment $B\vartheta^\prime=0$ along the elastica (also see Fig. \ref{fig:geom}). 
It also follows that 
\begin{equation}\label{eq:ODEs-cc}
\begin{cases}
{\vartheta^\prime}/(2 \lambda)   =  +\sqrt{c^2-\sin^2 \vartheta/2}~~{\rm for}~~ s \in [0,\ell^*]
\\
\\
{\vartheta^\prime}/(2 \lambda)   =  - \sqrt{c^2-\sin^2 \vartheta/2}~~{\rm for}~~ s \in (\ell^*,\ell/2]\\
\end{cases}.
\end{equation}
The change of the variables
\begin{equation}\label{eq:cov}
\sin \frac{\vartheta}{2} = c~ \sin \phi,~~~{\rm and}~~~{\rm d} \vartheta = \frac{2 c \cos \phi {\rm d} \phi}{\sqrt{1- c^2 \sin^2 \phi}},
\end{equation}
is convenient for the integration of \eqref{eq:ODEs-cc}, which gives 
\begin{align}\label{eq:load-angle-cc}
\ell~\sqrt{\frac{F}{B}} & = 4 {\rm K}(c),
\end{align}
where ${\rm K}(c)$ is the complete elliptic integral of the first kind (see (\ref{eq:def_cei})).

The distance $\ell_x$ between the ends of the elastica along the $x$-axis  is
\begin{equation}\label{eq:angle-deformation-cc-1}
{\ell_x} =  2 \int_0^{\ell/2} \cos \vartheta(s)~{\rm d} s,
\end{equation}
where  the integration was taken to  the midpoint $s=\ell/2$, because the Euler-type post-buckling displacement is symmetric with respect to the midpoint (see Fig. \ref{fig:geom}).  On the other hand, we can write 
\begin{equation}
\frac{\ell_x }{\ell}= 1 - \Delta \chi - \chi_{\rm cr}, 
\end{equation}
where we express the compression of the beam as $(u_--u_+)/\ell = \chi_{\rm cr} + \Delta \chi$.
Evaluating the integral in Eq. \eqref{eq:angle-deformation-cc-1} 
and using the equation  \eqref{eq:ODEs-cc} we deduce the required result
\begin{equation}\label{eq:angle-deformation-cc-2}
\lambda \ell_x = 4 \left[ 2 {\rm E}(c) - K(c)\right].
\end{equation}

 \subsubsection{Pre- and post-buckling approximations} \label{Sec232}
Here we use the extensible model of an elastic beam in the pre-buckling regime, and the classical Euler elastica inextensible model at the post-buckling stage, as in   equation (\ref{eq:load-displacement}).

 The Taylor expansion of the complete elliptic integrals around $c = 0$  is  
 \begin{equation}\label{eq:def_taylor_cei}
 {\rm  K}(c) = \frac{\pi}{2}\left( 1 + \frac{c^2}{4}  +  \frac{9}{64} c^4+ O(c^6) \right),~~~ {\rm  E}(c) = \frac{\pi}{2}\left( 1 - \frac{c^2}{4}  -  \frac{3}{64} c^4+ O(c^6) \right).
 \end{equation}
 Using the second equation in  \eqref{eq:load-displacement} and assuming $ 0< c \ll 1$ 
 we get 
 \begin{equation}\label{eq:cchi}
 c^2 =  {\Delta \chi}  - \frac{1}{8} (\Delta \chi)^2+ O((\Delta \chi)^3).
 \end{equation}
Using Eq. \eqref{eq:cchi} and the first equation in 
\eqref{eq:load-displacement}, we deduce the load-displacement relation 
\begin{equation}\label{eq:load-displacement2}
\ell  \sqrt{\frac{{\cal F}}{B}}= 2 \pi \left( 1 + \frac{\Delta \chi}{4}+\frac{7}{64} (\Delta \chi)^2 + O( (\Delta \chi)^3) \right).
\end{equation}

In the limit case when $\Delta \chi=0$, this implies ${\cal F}={P_c}= 4 \pi^2 B / \ell^2$, which is the first critical buckling load $P_c$ for a beam with the clamped ends. 
Hence, for small $\Delta \chi$ the function ${ F}(\chi)$ can be expressed as follows
\begin{equation}\label{eq:se-interaction}
{ F}(\chi) = P_c \left( 1 + \frac{\Delta \chi}{2}+\frac{9}{32} \Delta \chi^2 + O[ \Delta \chi^3] \right).
\end{equation}
\color{black}
In this pre-buckling regime, the  interaction between the masses  is linear with respect to the relative displacement, 
\emph{i.e.} in Eq. \eqref{eq:bc-derivative} we use 
   \begin{equation}\label{eq:f-prebuck}
   { F}(\chi) = E_1 S  \chi ,~~~{\rm for}~~\chi < \chi_{\rm cr},
   \end{equation} 
where we have chosen a positive sign for compression (\emph{i.e.} positive $\cal F$ for positive $\chi$). 
The  notation $\chi_{\rm cr}$ is used for the critical 
axial compression level
\begin{equation} 
\chi_{\rm cr} = \frac{P_c}{E_1 S} = 4 \pi^2 \frac{r^2}{\ell^2},~~~{\rm with}~~~ r^2 = \frac{\cal I}{S}.
\end{equation}
 \section{Time-harmonic scattering in the pre-buckling regime\label{sec:th-scattering}}
 Here, we focus on the pre-buckling regime where the interaction between the masses is given by Eq. \eqref{eq:f-prebuck}. We seek a solution to 
 \eqref{eq:pde-longitudinal}, with boundary conditions   \eqref{eq:bc-derivative} and \eqref{eq:bc-continuity},  
 in the form 
    \begin{equation}\label{eq:l-pde-longitudinal-solution}
u(x,t) = \begin{cases}  {\cal T} e^{i(k(x-\ell/2)-\omega t)}, & \mbox{for } x\ge \ell/2 \\ u_0~e^{ik(x+\ell/2)-i\omega t}+{\cal R} e^{-ik(x+\ell/2)-i\omega t} , & \mbox{for } x\le-\ell/2  \end{cases}, 
  \end{equation}
   where $u_0/\ell\ll\chi_{cr}$,  $k=\omega/v$ is the wave number, $\omega$ is the angular frequency, and ${\cal R}$ and ${\cal T}$ are  the reflection and transmission coefficients (to be determined). 
 Using Eqs \eqref{eq:bc-derivative} and  \eqref{eq:l-pde-longitudinal-solution} we deduce
 \begin{align}\label{eq:l-upum}
m_-\frac{{\rm d}^2u_-}{{\rm d} t^2}&=-{E_1 S}~(u_--u_+)/\ell + ikES\left(u_0-{\cal R}\right)e^{-i\omega t}, \nonumber\\
m_+\frac{{\rm d}^2u_+}{{\rm d} t^2}&={E_1 S}~(u_--u_+)/\ell  - ikES{\cal T}e^{-i\omega t}.
 \end{align}
 The continuity conditions \eqref{eq:bc-continuity} impose 
 \begin{equation}\label{eq:bc-continuity-1}
 u_- = \left(u_0+{\cal R}\right)e^{-i\omega t},~~~u_+= {\cal T} e^{-i\omega t}.
 \end{equation}
 Substitution of Eqs \eqref{eq:bc-continuity-1} into  \eqref{eq:l-upum} gives 
 \begin{align}\label{eq:R&T-system}
 -m_-\omega^2(u_0+{\cal R})&=E_1 S \frac{{\cal T}-{\cal R}-u_0}{\ell} + ikES\left(u_0-{\cal R}\right), \nonumber \\
-m_+\omega^2 {\cal T}&=-E_1 S\frac{{\cal T}-{\cal R}-u_0}{\ell}  - ikES{\cal T},
 \end{align}
 which is a linear algebraic system of equations for the scattering coefficients ${\cal R}$ and ${\cal T}$. The system can be  recast into the matrix form
 \begin{equation}\label{eq:R&T-matrix-form-1}
 \hat{\cal M}{\bf a}={\bf b}, 
 \end{equation}
 where ${\bf a}^{\rm T}=({\cal R}, {\cal T})$,   
 \begin{small}
 \begin{align}\label{eq:matrix_defs-1}
 \hat{\cal M} &= \begin{pmatrix}
 -m_- \omega^2 + 
 E_1 S/\ell + i \omega ES/ v & -E_1 S/\ell\\
 \\
-E_1 S/\ell & -m_+ \omega^2 + i \omega E S/ v + E_1 S/\ell
 \end{pmatrix},~
  \end{align}
\end{small}
 \noindent {\rm and} 
  \begin{small}
 \begin{align}
 {\bf b}&=\begin{pmatrix}
 m_- \omega^2 + i ES \omega / v - {E}_1 S/\ell \\
 \\
 E_1 S/\ell 
 \end{pmatrix}.
 \label{eq:matrix_defs-1a}
 \end{align}
\end{small}

The solution of  \eqref{eq:R&T-matrix-form-1} gives the reflection and transmission coefficients
\begin{equation}\label{eq:R&T}
{\cal R}=\frac{1}{{\cal D}} \left[ \frac{\omega}{\omega_0}-i \frac{\Delta m}{M} \frac{\omega^2}{\omega_-^2} -\frac{\omega \omega_0}{\omega_-^2}\left( 1-\frac{\omega^2}{\omega_+^2}\right) \right],~~~{\cal T}= -\frac{2 i}{\cal D}.
\end{equation}
Here 
 \begin{align}
 {\cal D}&= \frac{\omega}{\omega_0} + \frac{\omega \omega_0}{\omega_-^2}\left(1-\frac{\omega^2}{\omega_+^2}\right) - i \left( 2-\frac{\omega^2}{\omega_-^2}\right),\nonumber\\
 \omega_0&=\frac{v E_1}{\ell E}, ~~~\omega_+^2=\frac{E_1 S}{\mu\ell},~~~ \omega_-^2=\frac{E_1 S}{M\ell},~~~,\nonumber\\M&=m_++m_-, ~~~ \Delta m= m_+-m_-,~~~{\rm and}~~~\mu= \frac{m_-m_+}{m_++m_-}. 
 \end{align}
   \begin{figure}[h!]
 	\centering
 	\includegraphics[width=0.7\textwidth]{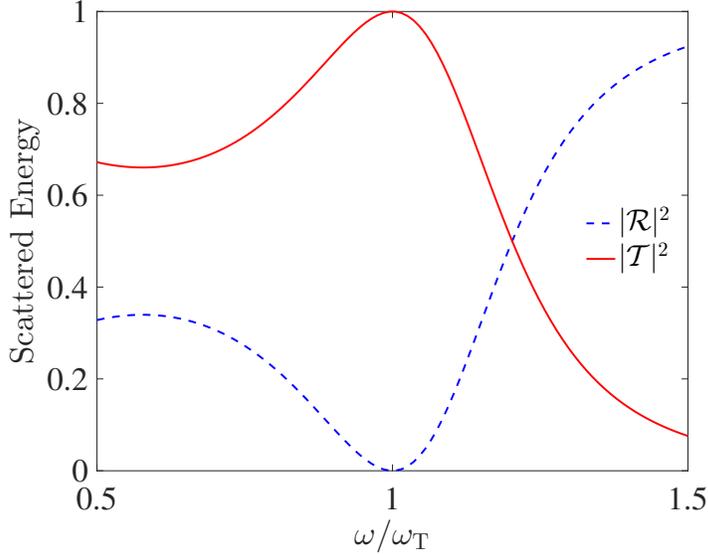}
 	\caption{\label{fig:R&T_linear} Reflectance and transmittance calculated using Eq. \eqref{eq:R&T}. 
	The transmission resonance occurs at $\omega=\omega_{\rm T}$ given in Eq. \eqref{eq:wT}. We have chosen the frequencies $\omega_0=3$, $\omega_+=2.45$, $\omega_-=1.23$, resulting in $\omega_{\rm T}=2.24$.}
 \end{figure}
 \subsection{Transmission resonance\label{sec:trans-res}}
The reflection coefficient in Eq. \eqref{eq:R&T} vanishes when 
\begin{equation}\label{eq:roots_Req0}
\frac{\omega_0}{\omega_-^2\omega_+^2}\omega^2-i \frac{\Delta m}{M} \frac{\omega}{\omega_-^2} + \frac{1}{\omega_0}-\frac{\omega_0}{\omega_-^2}=0 
\end{equation}
and the corresponding roots are
\begin{equation}\label{eq:roots_Req0a}
\omega=\frac{\omega_+^2}{\omega_0}\left\{ \frac{i}{2} \frac{\Delta m}{M} \pm \left[ -\frac{1}{4}\frac{\Delta m^2}{M^2} +\frac{1}{\omega^2_+}\left(\omega_0^2-\omega_-^2\right) \right]^{1/2}\right\}.
\end{equation}
Eq.  \eqref{eq:roots_Req0}  has real roots  if 
\begin{equation}\label{eq:cond-mass}
\Delta m =0,~~~{\rm and}~~~\omega_0^2>\omega_-^2.
\end{equation}
The radian frequency corresponding to the transmission resonance is 
\begin{equation}\label{eq:wT}
\omega = \omega_{\rm T} = \frac{\omega_+^2}{\omega_0} \left[ \frac{\omega_0^2-\omega_-^2}{\omega_+^2}\right]^{1/2}.
\end{equation}
If we assume that $E_1=E$, then Eq. \eqref{eq:wT} can be rewritten as 
\begin{equation}\label{eq:wT-1}
\omega_{\rm T} =  \frac{E S}{m v}\left[ \frac{\bar{\rho}}{\rho}-1\right]^{1/2},~~~{\rm where}~~~\bar{\rho}=\frac{2 m}{\ell S},
\end{equation}
and 
$m=m_-=m_+$.
  \section{ Transient  scattering from the non-linear interface} 
  \label{transient}
We note that the time-harmonic form of the displacement field \eqref{eq:l-pde-longitudinal-solution} in the 
problem with the non-linear interface is no longer valid.  At both sides from the non-linear interface, a general solution of the wave equation is employed and it represents traveling waves in the semi-infinite rods whose shape is determined as a result of the interaction with the non-linear interface. A solution to the transient  problem \eqref{eq:pde-longitudinal} with non-linear transmission conditions \eqref{eq:bc-derivative}  can be sought using the travelling wave Ansatz   

\begin{equation}\label{eq:nl-scattering-solution}
 u(x,t) = \begin{cases} u_1(x,t) =  u_0~{\rm cos}\left(\frac{\omega}{v}(x+\frac{\ell}{2}) -\omega t\right)+R\left(-\frac{\omega}{v}(x+\frac{\ell}{2}) -\omega t\right), & \mbox{if } x\le - \ell/2 \\ u_2(x,t) = T\left(\frac{\omega}{v}(x-\frac{\ell}{2}) -\omega t\right), & \mbox{if } x\ge\ell/2 \end{cases},
  \end{equation}
     \begin{figure}[t!]
\centering
\includegraphics[width=1.0\textwidth]{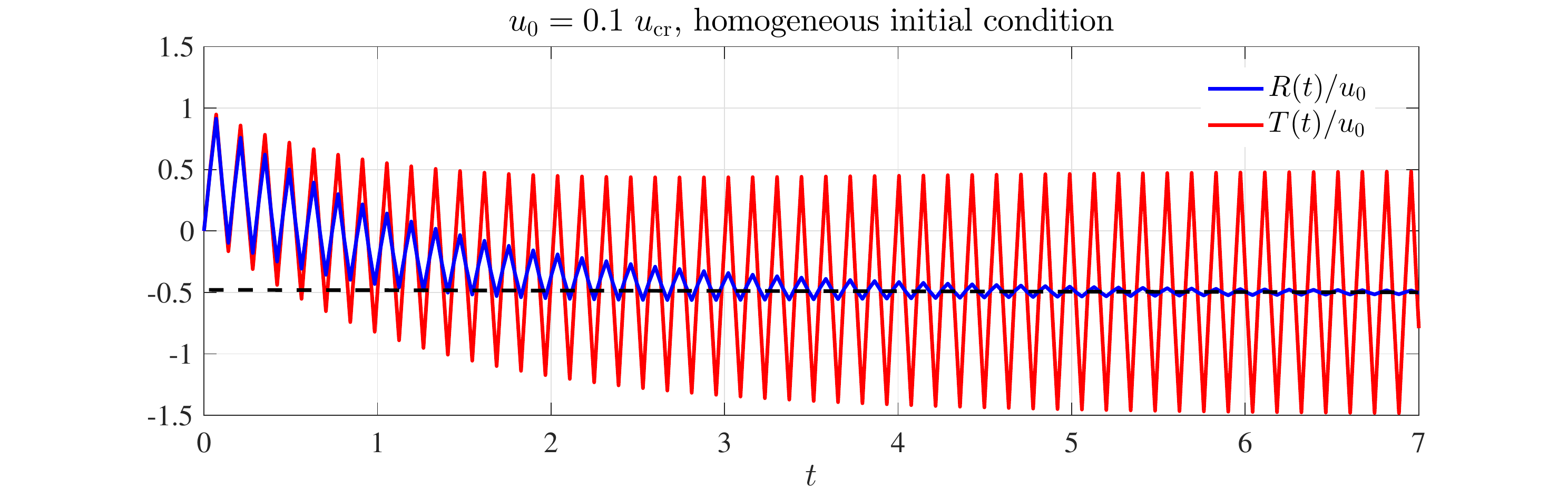}\hfill
(a)

\vspace{.1in}

\includegraphics[width=1.0\textwidth]{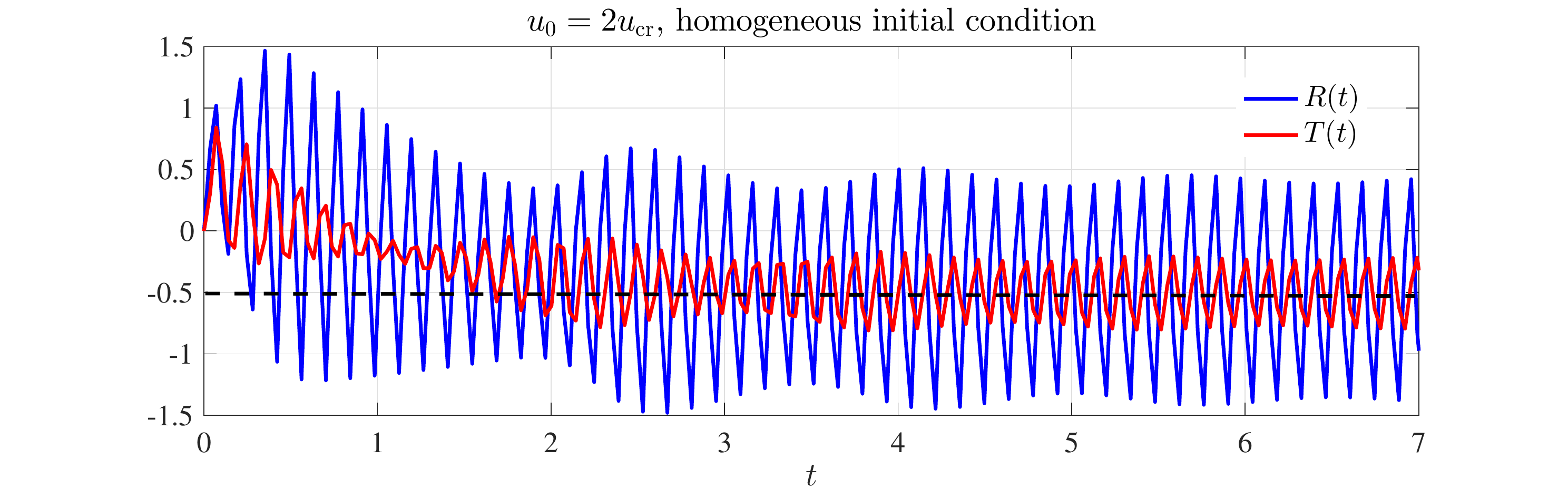}\\
(b)
\caption{\label{fig:RT-transient-homo-con} Transient solutions of the system 
\eqref{eq:system-odes} with homogeneous initial conditions \eqref{eq:homo-con}. The parameters are the same as in Fig. \ref{fig:f-nl}. The different panels refer to different values of the incident wave amplitude $u_0$. The frequency of the incident wave is chosen to be $\omega=\omega_{\rm T}=44.7$, for which a transmission resonance (see Fig. \ref{fig:R&T_linear}) in the time-harmonic linear problem occurs, as demonstrated in the diagram (a) for  $u_0=0.1~u_{\rm cr}$. 
For larger amplitude, the simulation in the diagram (b) shows a higher reflection. }
\end{figure}
   \begin{figure}[t!]
\centering
\includegraphics[width=1.0\textwidth]{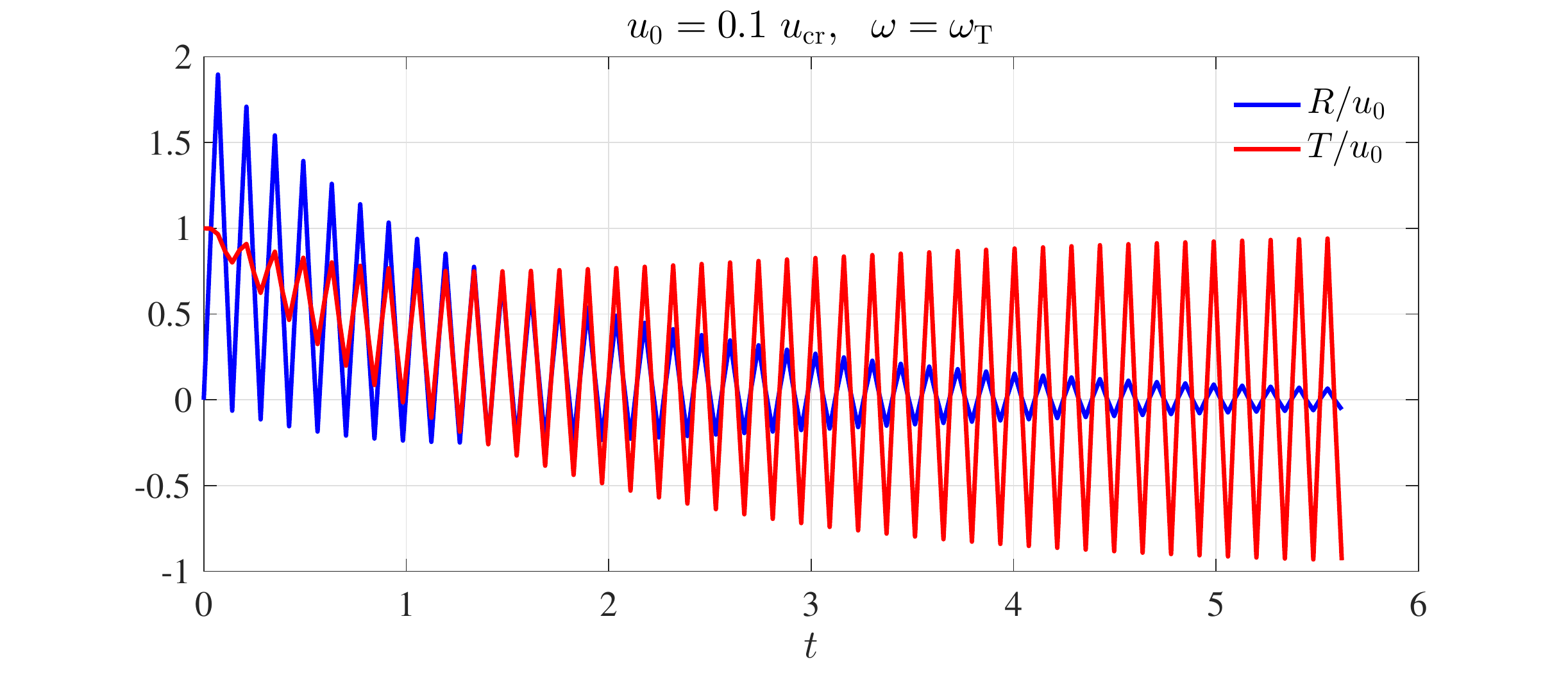}\hfill

(a)

\vspace{.1in}
\includegraphics[width=1.0\textwidth]{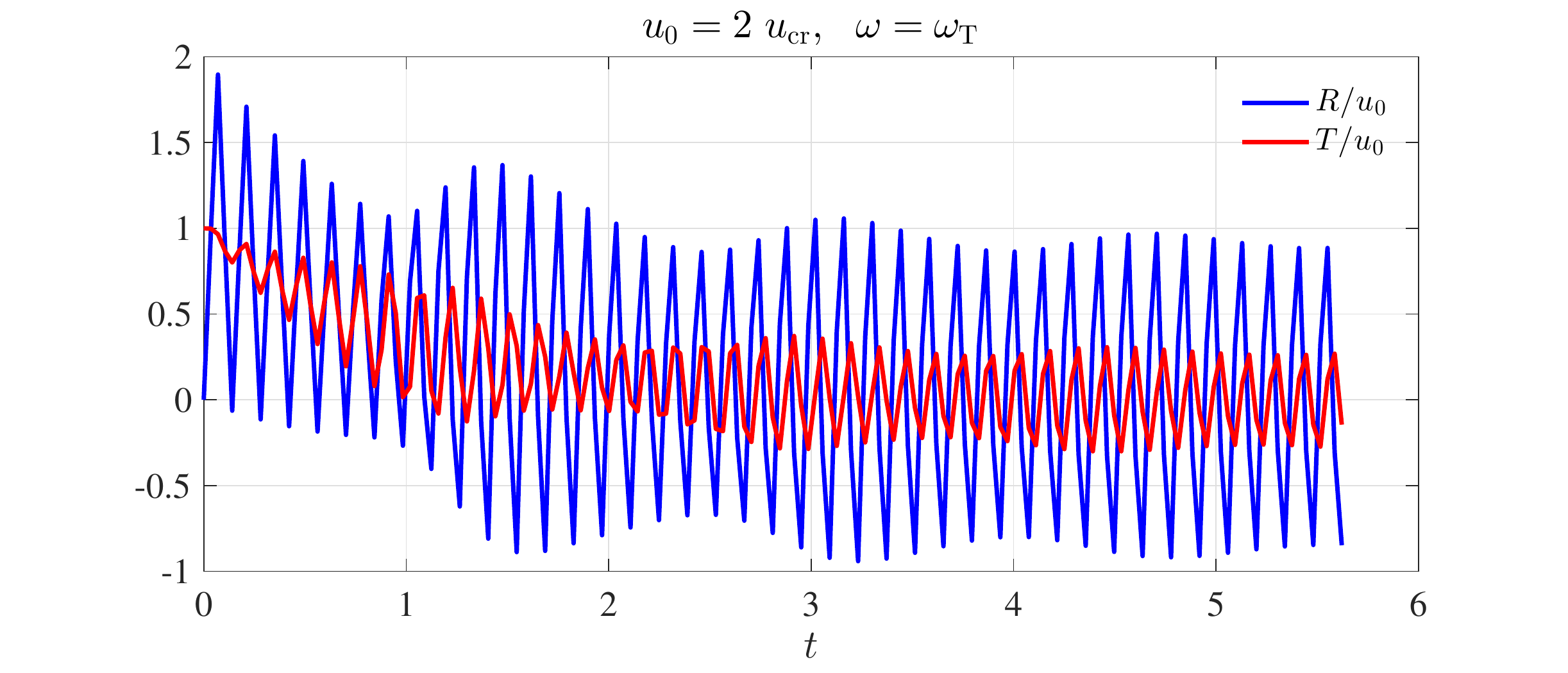}\\
(b)
\caption{\label{fig:RT-transient-non-homo-con} Transient solutions of \eqref{eq:system-odes} with parameters as in Fig. \ref{fig:f-nl} with non-homogeneous initial conditions \eqref{eq:non-homo-con}. The panels (a) and (b) refer to different values of the incident wave amplitude $u_0$. The frequency of the incident wave is chosen to be $\omega=\omega_{\rm T}=44.7$, for which a transmission resonance in the time-harmonic linear problem occurs.}
\end{figure}
  where we have chosen to express the functions $R$ and $T$ in the form of left-propagating and right-propagating waves, respectively. By introducing the variables $\tau = -\omega t$ and $\xi_\pm=\omega(\pm x/v - t)$, 
  we observe that 
  \begin{equation} \label{eq:diff-rules}
  \frac{\partial}{\partial x} f (\xi_{\pm} )= \mp \frac{\omega}{v}f^\prime(\xi_\pm),~~~{\rm and}~~~ \frac{\partial}{\partial t} f(\xi_\pm)= \omega f^\prime(\xi_\pm),
  \end{equation}
  where $f$ is a given scalar function. 
 Using \eqref{eq:nl-scattering-solution} and 
 \eqref{eq:diff-rules} 
 the transmission conditions \eqref{eq:bc-derivative} and \eqref{eq:bc-continuity} can be written in the form
  \begin{align}\label{eq:def-us}
  \left.\ddot{u}_1(x,t)\right|_{x=-\ell/2}&=-\omega^2\left(u_0 \cos(\tau)-R^{\prime\prime}(\tau)\right),\nonumber\\
 \left. \frac{\partial}{\partial x}u_1(x,t)\right|_{x=-\ell/2}&= \frac{\omega}{v}\left( u_0 \sin(\tau)  +R^{\prime}(\tau)\right),\nonumber\\
 \left. \ddot{u}_2(x,t) \right|_{x=+\ell/2}&=\omega^2T^{\prime\prime}(\tau),\nonumber\\
 \left.\frac{\partial}{\partial x}u_2(x,t)\right|_{x=+\ell/2}&=-\frac{\omega}{v} T^\prime(\tau).
  \end{align}
 Furthermore, the transmission conditions   \eqref{eq:bc-derivative} yield 
  \begin{align}\label{eq:bc-derivative-nl}
  -\omega^2m_-\left(u_0 \cos(\tau)-R^{\prime\prime}(\tau)\right)&=\frac{ES \omega}{v}\left( u_0 \sin(\tau)  +R^{\prime}(\tau)\right)-{\cal F}(u_0 \cos(\tau) + R(\tau), T(\tau)),\color{black}\nonumber\\
 \omega^2 m_+ T^{\prime\prime}(\tau) &= \frac{ E S \omega}{ v} T^\prime(\tau) + {\cal F}(u_0 \cos(\tau) + R(\tau), T(\tau)),
  \end{align}
  or equivalently 
   \begin{align}\label{eq:system-odes}
R^{\prime\prime}(\tau)&= u_0 \cos(\tau) + \frac{E S}{m_-\omega v} u_0 \sin(\tau)  + \frac{ES}{m_- \omega v} R^\prime(\tau)  - \frac{{\cal F}(u_0\cos(\tau)+ R(\tau),T(\tau))}{m_-\omega^2},\nonumber\\
T^{\prime\prime}(\tau) &= \frac{ E S}{v \omega m_+} T^\prime(\tau) + \frac{{\cal F}(u_0\cos(\tau)+ R(\tau),T(\tau))}{m_+\omega^2}.
  \end{align}
  In Eqs \eqref{eq:system-odes}, we use the expression \eqref{eq:load-displacement} for the load-displacement relation ${\cal F}(u_-,u_+)$, with notation consistent with the one already introduced in the context of Eq. \eqref{eq:bc-derivative}.
     \begin{figure}[t!]
\centering
\includegraphics[width=0.45\textwidth]{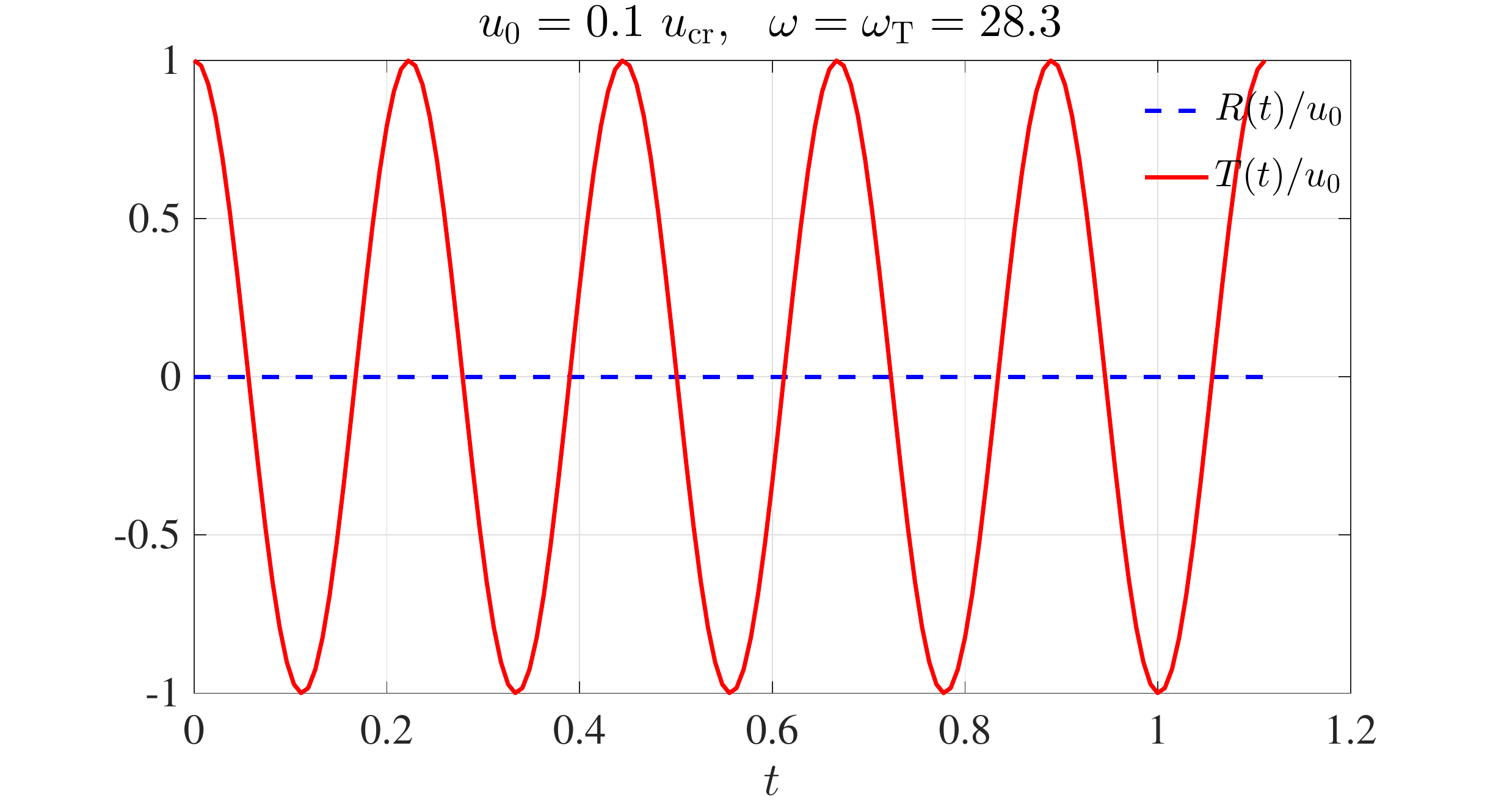}\hfill
\includegraphics[width=0.52\textwidth]{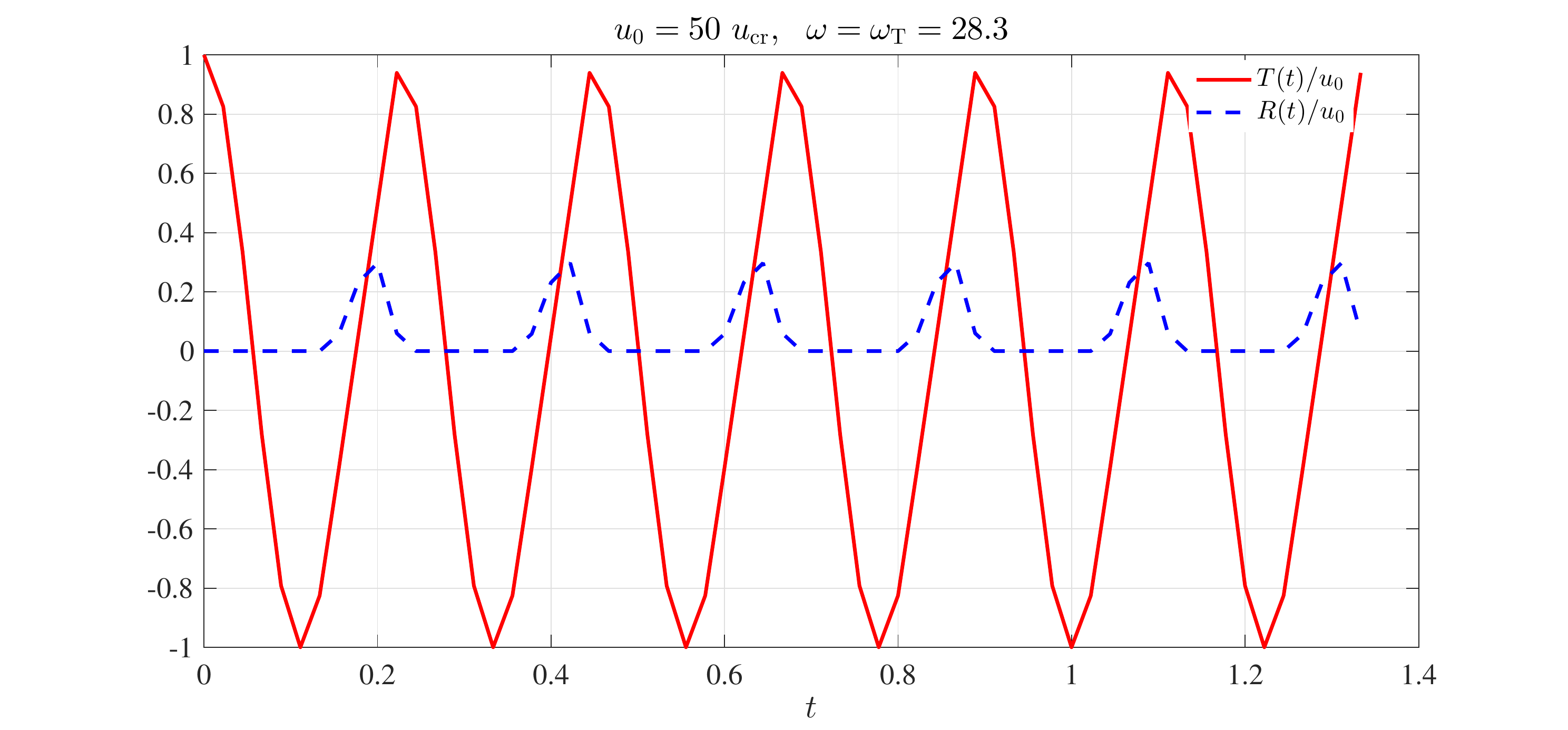}\\
(a) ~~~~~~~~~~~~~~~~~~~~~~~~~~~~~~~~~~~~~~~~~~~~~~~~~~~~~(b)
\caption{\label{fig:RT-transient-1} Transient solutions of the system  \eqref{eq:system-odes} with parameters as in Fig. \ref{fig:f-nl} and non-homogeneous initial conditions \eqref{eq:non-homo-con}. The different panels refer to different values of the incident wave amplitude $u_0$. The frequency of the incident wave is chosen to be $\omega=\omega_{\rm T}=28.3$, for which a transmission resonance in the time-harmonic linear problem exists (see $u_0=0.1~u_{cr}$ and \emph{cf.} with Fig. \ref{fig:R&T_linear}).}
\end{figure}
\paragraph{Initial conditions and numerical simulations.} In the computations discussed here, $R(t)$ and $T(t)$ are measured at the left and right ends of the interface, respectively. In order to compute functions $R$ and $T$, a set of initial conditions for the system \eqref{eq:system-odes} has to be chosen. 

First, we consider homogeneous initial conditions 
\begin{equation}\label{eq:homo-con}
R(0)=T(0)=R^\prime(0)=T^\prime(0)=0.
\end{equation}
For this case, the results of the numerical simulations are shown in Fig. \ref{fig:RT-transient-homo-con}, which includes two parts: in the first part  (a) the amplitude of the incident wave is small, while in the second part (b) the amplitude has been increased. In both cases we used the transmission resonance frequency for the incident wave. 

Second, the time-harmonic analysis conducted in section \ref{sec:trans-res}, suggests the use of initial conditions compatible with the prediction of the time-harmonic analysis.  Thus, at the transmission resonance frequency, we choose 
\begin{equation}\label{eq:non-homo-con}
R(0)=R^\prime(0)=T^\prime(0)=0,~~~{\rm and}~~~T(0)=u_0.
\end{equation}
For this case, the results of the simulations are shown in Fig. \ref{fig:RT-transient-non-homo-con}, which also includes two parts, corresponding to a small amplitude (part (a)) and a finite amplitude (part (b)) of the incident wave.

In both cases, it has been shown that the small amplitude  incident wave is consistent with the transmission resonance, while increase in the amplitude triggers the non-linear response from the structured interface.

In both cases it has been observed that  the non-linear response of the structured interface has suppressed the resonance transmission, and hence has lead to a higher reflection from the interface region. 

Furthermore, in Fig. \ref{fig:RT-transient-1} we present the case of a lower frequency for the transmission resonance. The initial conditions (\ref{eq:non-homo-con}) are used in this simulation. The part (a) of the figure is fully consistent with the transmission resonance and shows very low reflection, while the part (b) corresponding to a larger amplitude shows a completely different process, where the reflection from the non-linear interface is significant. In this frequency regime, the function $R(t)$ is asymmetric which corresponds to relatively low resistance of the buckled beam when the left end of the interface moves in the positive direction. 

\section{Concluding remarks and discussion}
\label{conclusion}

Motivated by the work of Bigoni {\em et al.} \cite{Bigoni2012_PRSA,Bigoni2014_JMPS, Bigoni2015_MM, Bigoni2014_PRSA, bigoni2012nonlinear}, and Maurin and Spadoni \cite{Maurin2014_WM,Maurin2014_JSV,Maurin2016_WM1,Maurin2016_WM2} we have identified interesting transmission/reflection configurations in the context of the dynamic response of a non-linear structured interface embedded into the linear elastic bar. It has been demonstrated that a transmission resonance, existing in the linearised system, can be suppressed via the non-linear response of the interface interacting with an incident transient waveform.    

This study brought further thoughts, based on the homogenisation of a nonlinear chain. Assuming a linear semi-infinite elastic bar,   
 \begin{equation}\label{eq:pde-}
 \frac{\partial^2 u}{\partial t^2}- c^2  \frac{\partial^2 u}{\partial x^2}=0,~~~x<0,
 \end{equation}
is in contact with a semi-infinite non-linear chain of point-masses of magnitude $m$ which occupies the region $x>0$ (as in \cite{Maurin2014_WM,Maurin2014_JSV}), one may be interested in evaluating the reflection from the interface subjected to an incident pulse. In Eq.  \eqref{eq:pde-}, $u(x,t)$ is the longitudinal displacement,  and $c$ is the longitudinal wave speed. The semi-infinite non-linear chain of masses is assumed to possess a homogeneous buckling level $\ell_x$, and we consider the dynamics response of the system about such buckled configuration, in the post-buckling regime and for small amplitude with respect to such buckling level. The homogenisation approach was discussed in 
\cite{Maurin2016_WM1}, where the dynamic response was studied for a periodic  post-buckled lattice, with an elementary cell including a buckled beam with hinge junctions.  In the long wavelength regime and when the wave amplitude is smaller than the static buckling level $\ell_x$  
the homogenisation approximation is governed by the equation of the Boussinesq type \cite{Maurin2014_WM}
 \begin{equation}\label{eq:boussinesq}
\frac{\partial^2 u}{\partial t^2}= c_0^2  \frac{\partial^2 u}{\partial x^2} + 2 c_0 \gamma \frac{\partial^4 u}{\partial x^4} - c_1^2   \frac{\partial^2 u}{\partial x^2}   \frac{\partial u}{\partial x},~~~x>0,
 \end{equation}
where $c_0^2=P_c \ell_x^2/(2^7 m \ell)(1+\ell_x/\ell)^3$, $c_1^2 =3/2^{11} (P_c \ell^3_x / (\ell^2 m) (1+\ell_x/\ell)^4$ and $\gamma=c_0 \ell_x^2/24$; the notations of Section \ref{sec:nl-int} are used here, apart from $P_c$ which represents the critical buckling load for a beam with hinged ends. In the above equation, 
the coefficients $c_0$ and $c_1$ have the dimensions of the wave speed, and the coefficient $\gamma$ near the fourth-order term is determined via the higher-order lattice homogenisation procedure.

We note the different orders of the equations \eqref{eq:pde-}, \eqref{eq:boussinesq} at different sides of the interface, which suggests the requirement for the boundary layer analysis in the transition region. Problems of this kind also occur for formulations based on configurational force models involving a dynamic transition from the longitudinal elastic waves to buckled beams across the sliding sleeve interface. 

The dynamics of the non-linear interface incorporating a system of movable clamps and buckled beams can be traced without additional use of the boundary layers near the junction region; in this case,  the shape of the envelope function may depend on the initial conditions and the amplitude of the incident wave. 
The appropriate choice of initial conditions may lead to the higher transmission over the given time interval. Although this observation is not surprising, it is important as generally transmission resonances observed in linearised time-harmonic formulations for structured interfaces are suppressed by the non-linear interaction in the transient regime (see Figs \ref{fig:RT-transient-homo-con} and \ref{fig:RT-transient-non-homo-con}).   In addition, by observing Fig. \ref{fig:RT-transient-homo-con} we note that, at sufficiently large times, the reflection functions converge to a constant non-zero asymptotic value (see dashed black line), while in Fig. \ref{fig:RT-transient-non-homo-con} the average value of the reflection function converges to zero. This corresponds to the two-mass system moving its centre of mass over time as a result of the driving elastic waves. The asymptote is reached when the momentum per unit time provided by the wave is equilibrated by the surrounding elastic bar.  A non-zero average asymptotic value could be measured experimentally, by measuring the time-dependent phase shift of outgoing waves. 

Further extension of the analysis of dynamic non-linear interfaces is envisaged to configurational force models, which may incorporate an additional boundary layer study to account for possible discontinuity in the displacements or their derivatives.

\section{Acknowledgements}

Part of this work has been conducted while DT was a post-doctoral fellow at INSA-Lyon, France;  funding from the French Research Agency ANR Grant ``METASMART'' (ANR-17CE08-0006) is gratefully acknowledged. ABM and NVM acknowledge the support of the UK EPSRC Program Grant EP/L024926/1.  A part of this paper was initiated while DT and ABM were visiting the University of Trento, Italy; the support of the ERC Advanced Grant ``Instabilities and nonlocal multiscale modelling of materials" ERC-2013-ADG-340561-INSTABILITIES is gratefully acknowledged.

\end{document}